\documentclass{aa}

\usepackage[varg]{txfonts}
\usepackage{natbib}
\usepackage{graphicx}
\usepackage{times}
\usepackage{color}
\usepackage{hyperref}
\hypersetup{
  hidelinks,
  colorlinks=true,
  linkcolor=blue,
  filecolor=blue,
  urlcolor=blue,
  citecolor=blue,
}
\usepackage{bm}
\usepackage{enumitem}   
\usepackage{scalerel}
\usepackage{multirow}
\usepackage{dblfloatfix}
\usepackage[flushleft]{threeparttable}

\bibpunct{(}{)}{;}{a}{}{,}
\usepackage{tabularx}
    \newcolumntype{C}{>{\centering\arraybackslash}X}
    \newcolumntype{L}{>{\raggedright\arraybackslash}X}
    \newcolumntype{R}{>{\raggedleft\arraybackslash}X}
\usepackage{array}

\newcommand{\flaree}[3]{$(#1 \pm #2) \times 10^{#3}$}

\usepackage{siunitx,booktabs}
\usepackage[version=4]{mhchem}
\usepackage{collcell}

\begin{document}
   \title{Observational signs of limited flare area variation and peak flare temperatures estimations in main-sequence flaring stars}

   \author{K. Bicz\inst{1} \and R. Falewicz\inst{1, 2} \and P. Heinzel\inst{2,3} \and P. Pre\'s\inst{1} \and D. Mo\'zdzierski\inst{1} \and A. Pigulski\inst{1} \and D. Marchev\inst{4} \and K. Kotysz\inst{1} \and T. Atanasova\inst{4} \and G. Yordanova\inst{4} \and A. Georgiev\inst{4}}

   \institute{Astronomical Institute, University of Wroc\l{}aw, Kopernika 11, 51-622 Wroc\l{}aw, Poland
              \and
             University of Wroc\l{}aw, Centre of Scientific Excellence - Solar and Stellar Activity, Kopernika 11, 51-622 Wroc\l{}aw, Poland\and
             Astronomical Institute, Academy of Sciences of the Czech Republic, 25165 Ond\v{r}ejov, Czech Republic\and
             Department of Physics and Astronomy, Shumen University "Episkop Konstantin Preslavski", 115 Universitetska Str., 9700 Shumen, Bulgaria\\
             \email{bicz@astro.uni.wroc.pl}
             }
 
  \abstract{In the study of stellar flares, traditional method of calculating total energy emitted in the continuum assumes the emission originating from a narrow chromospheric condensation region with a constant temperature of 10$\,$000$\,$K and variable flare area. However, based on multicolor data from 7 new flares observed in Białków and Shumen observatory and 8 flares from \cite{Howard_2020} observed on 10 main-sequence stars (spectral types M5.5V to K5V -- 9 M-dwarfs and 1 K-dwarf) we show that flare areas had a relative change in the range of $10\% - 61\%$ (for more than half of the flares this value did not exceed 30\%)
throughout the events except for the impulsive phase, and had values starting from $50 \pm 30\,$ppm to $300 \pm 150\,$ppm for our new flares and from $380\pm200\,$ppm to $7\,600 \pm 3\,000\,$ppm from \cite{Howard_2020} data, while their temperature increased on average by the factor 2.5. 
The peak flare temperatures for our seven observed flares ranged from $5\,700 \pm 450\,$K to $17\,500 \pm 10\,050\,$K. Five of these flares had their temperatures estimated using the Johnson-Kron-Cousins B filter alongside {\it TESS} (Transiting Exoplanet Survey Satellite) data, one flare was analyzed using the SLOAN $g'$ and $r'$ bandpasses, and another was evaluated using both the SLOAN $g'$ and $r'$ bandpasses and {\it TESS} data.
Using flare temperature and area from our data and data from \cite{Howard_2020}, along with the physical parameters of stars where the flares occurred, we developed a semi-empirical grid that correlates a star's effective temperature and flare amplitude in {\it TESS} data with the flare's peak temperature. This allows interpolation of a flare's peak temperature based on the star's effective temperature (ranging from $2\,700\,$K to $4\,600\,$K) and flare amplitude from {\it TESS} observations. Applying this grid to $42\,257$ flares from {\it TESS} survey, we estimated peak flare temperatures between $5\,700\,$K and $38\,300\,$K, with most flares showing peak black-body temperatures around $11\,100 \pm 2\,400\,$K. }


   \maketitle
    \titlerunning{Flare Areas and Temperature in main-sequence Flaring Stars}

\section{Introduction}\label{sec:introduction}

Flares are random and unpredictable events on the surfaces of stars, characterized by the intense release of magnetic energy in the outer stellar atmosphere. These flares produce a range of observable effects across various wavelengths, emitting energy in both emission lines and the continuum. The continuum, especially from the far-ultraviolet (FUV) to the optical range, dominates the energy output. During the peak of most flares, a few percent of the total energy is in the emission lines, which can rise to 20\% during the gradual decay phase \citep{Kowalski_2013}. In certain cases, line emission can contribute up to 50\% of the total flare energy \citep{Hawley_2007}.

On the Sun, our ability to see the detailed surface observations allows us to study flares frequently and in great detail, providing significant insights into flare processes and their consequences. The total energy emitted by solar nanoflares and microflares, observed in X-ray on very short timescales, can be as low as $10^{21} - 10^{24}\,$erg (impulsive events occur at a frequency of $\sim$25 events per minute with a typicalpp lifetime of $\sim$10 minutes) \citep{Pres_2005, Upendran_2022}. However, this energy can reach levels typical of a solar flares visible in white light, ranging from $10^{29}\,$erg to approximately $10^{32}$-$10^{33}\,$erg \citep{Shibata_2002, Cliver_2013}, with durations spanning from several minutes to several hours. To date, no solar flare having energy exceeding 10$^{33}\,$erg, have been observed. For other active stars, although it is not possible to observe flares with the same level of detail, we are still able, under favorable conditions, to estimate their location on the surface of the star \citep{Ilin_2021, Bicz_2024}. Stellar flares can be over a thousand times more intense than solar flares, with durations ranging from minutes to several hours \citep{Pietras_2022}. In extraordinary cases, a flare can last over a day \citep{Bicz_2024}.

The continuum emission of stellar flares is often modeled as black-body radiation with a fixed temperature of 10$\,$000$\,$K, while the flare's area changes over time \citep{Shibayama_2013, Namekata_2017, Gunther_2020, Bicz_2022, Pietras_2022}. In this model, the flare area is described as a unified radiating structure, where white-light ribbons and cool loops are not distinct entities but interconnected components of a single composite system on the star’s surface.
\begin{figure*}[ht!]
    \centering
    \includegraphics[width=\textwidth]{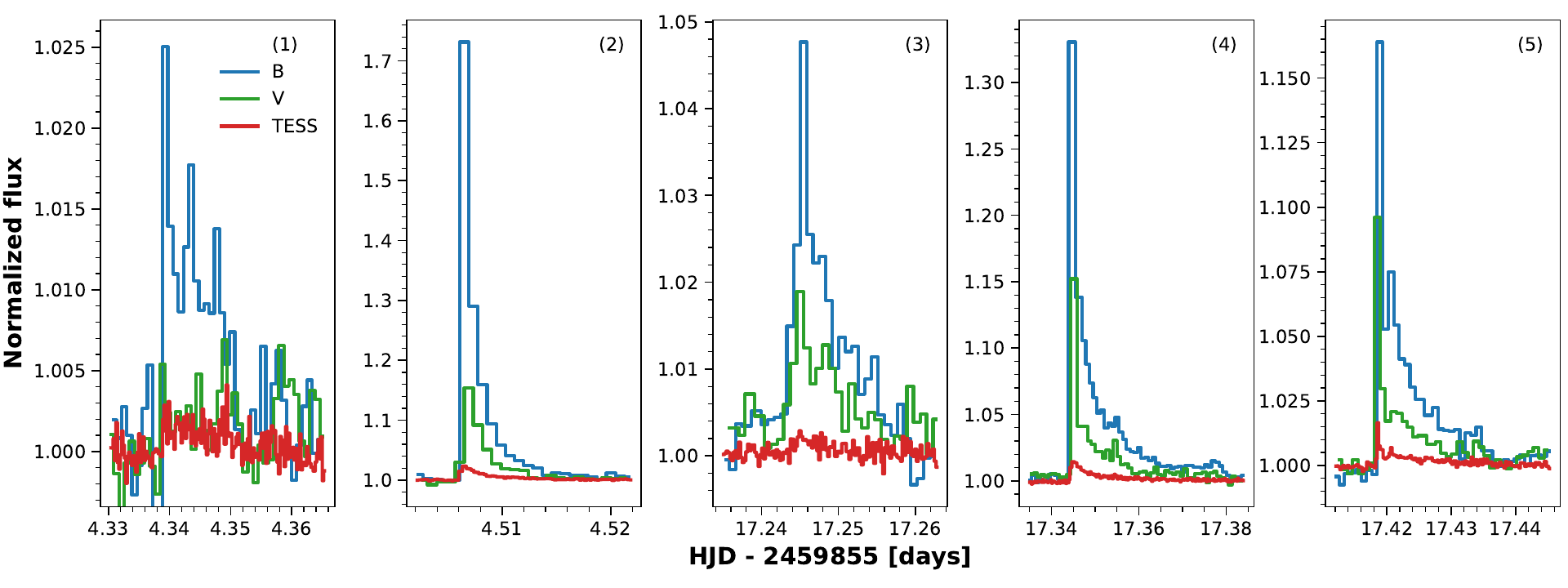}
    \caption{Normalized light curves of five flares of EV Lac in three different passbands. The red, blue, and gree curves represent the B-band, V-band made in Bia\l{}k\'ow Observatory, and the 20$\,$s TESS observations, respectively. The panels are labelled with the flare number (Table \ref{tab:obsbialshu}).}
    \label{fig:bialkow}
\end{figure*}
It is important to note that, during a flare, temperature evolves over time similarly to solar observations, as shown in studies by \cite{Gershberg_1971, Kunkel_1975, Mochnacki_1980}, and others. \cite{Hawley_1992} modeled atmospheric responses to flare energy, finding that conductive heating balanced optically thin cooling in the transition region. However, discrepancies in emission ratios and line profiles indicated additional heating in the upper chromosphere. Observed flare continuum spectra, resembling blackbody radiation ($8500$-$9500\,$K), were attributed to photospheric reprocessing of upper chromospheric UV-EUV emissions.
\cite{Hawley_1995} analyzed optical and coronal data, finding strong evidence of a stellar "Neupert effect," which supports chromospheric evaporation models.  They further argued that the spatial correlation between white-light and hard X-ray emissions in solar flares, along with the identification of hard X-rays as nonthermal bremsstrahlung from accelerated electrons, indicates that flare heating on dMe stars follows the same electron precipitation mechanism observed on the Sun. Their study of a large EUV flare emphasized the role of coronal loop length in flare energetics, further extending the solar flare model to stellar environments.
\cite{Zhilyaev_2007} studied EV Lac flares, confirming high-frequency oscillations tied to hydrogen plasma state transitions. Peaks matched blackbody temperatures of $17\,000$-$22\,000\,$K, with flares covering $\sim\!$1\% of the stellar disk.
\cite{Kowalski_2013} explored M-dwarf flares, showing that electron beams primarily heated the chromosphere. Numerical models highlighted wave interactions and localized heating, predicting NUV and optical continua at temperatures $\geq12\,000\,$K, consistent with observed flare properties.
The total energy of the flare is determined by its black-body temperature and the evolving flare area. Although, the actual black-body temperatures of superflares are still uncertain. For M-dwarf flares, continuum temperatures typically range between 9$\,$000 and 14$\,$000$\,$K \citep{Kowalski_2013}, but in some cases, they can surpass 40$\,$000$\,$K \citep{Robinson_2005, Kowalski_Allred_2018, Froning_2019}.
Substantial temperature variations are observed during stellar flares as the energy source propagates from the lower chromosphere into the coronal regions \citep{Livshits_1981, Fisher_1985, Fisher_1985b, Fisher_1985c, Fisher_1989}. Such a process is called chromospheric evaporation. During a solar or stellar flare near a magnetic reconnection area, local plasma escapes as a two-directional stream along the magnetic flux rope. The escaping plasma is heated, and electrons are accelerated to non-thermal energies. Both thermal conduction and the accelerated electrons can deliver a high energy flux to the dense chromosphere, that responds with explosive heating and expansion. As a result, hot plasma flows upward and quickly fills the newly formed flare loops \citep{Fisher_1985, Falewicz_2014, Dudik_2016, Yang_2021}.

In this study, we measured the evolution of black-body temperatures and areas for seven flares using multi-color photometry. Five of these flares were observed at by Bia\l{}k\'ow Observatory (Poland) at  approximately one-minute intervals and the Transiting Exoplanet Survey Satellite ({\it TESS}; \cite{Ricker_2014}) at both two-minute and 20-second intervals. The remaining two flares were observed by {\it TESS} and the Shumen Observatory (Bulgaria). Eight additional flares from \cite{Howard_2020} were incorporated to enhance the statistical analysis. By utilizing the black-body temperature evolution data of these flares, we calculated the flare area evolution for eight additional events. 

Because multi-color observations of stellar flares are limited we tried to prepare an improved procedure that allows for the analysis of stellar flares using single-channel observations, comparable to those conducted with multiple channels. Using calculated temperature and derived area evolution of the analyzed flares we were able to create a grid allowing to interpolate the peak temperature of any flare that occurs on a main-sequence flaring star. To do so we need the amplitude of the flare calculated from {\it TESS} observations and the effective temperature of the star. 
This method works for the stars with effective temperatures in the range from 2$\,$700~K up to 4$\,$600$\,$K.

The paper is organized as follows. In Sect. \ref{sec:data}, we describe the simultaneous flare observations and host star properties. In Sect. \ref{sec:methods}, we describe our method for estimating the flare's temperature and area changes, as well as our calculations of the flare's energy. In Sect. \ref{sec:results}, we present the results and propose a new method for estimating the peak temperature of a flare. In Sect. \ref{sec:discussion}, we summarize and discuss the results.

\section{Observations}\label{sec:data}

\subsection{Bia\l{}k\'ow observations}
The observational data from Bia\l{}k\'ow Observatory of the University of Wroc\l{}aw, Poland were obtained using a 60~cm Cassegrain telescope equipped with an Andor Tech. DW432-BV back-illuminated CCD camera. This setup provides a field of view of $13' \times 12'$ in the B and V bandpasses of the Johnson-Kron-Cousins photometric system with approximately 70$\,$s cadence for each bandpass. Flare temperatures were estimated using the B/TESS flux ratio to avoid the effect of noise in the V-band photometry. To check if these temperatures were accurate enough, we compared the observed V-band signal with the signal expected from the calculated temperature and flare area. The results matched very well, with differences usually smaller than two standard deviations.

During the nights of October 6/7 and 19/20, 2022, five flares were observed on the active M-dwarf star EV Lacertae at Bia\l{}k\'ow Observatory. These flares have numbers from 1 to 5 in this paper (see Fig. \ref{fig:bialkow}). EV Lacertae (EV Lac) is classified as an M4Ve \citep{Lepine_2013} flaring star with an effective temperature of approximately $3\,300\,\pm\,200$~K, radius of $0.34\,\pm\,0.01\,$R$_\odot$, and mass of $0.32\,\pm\,0.02\,$M$_\odot$ (MAST\footnote{Barbara A. Mikulski Archive for Space Telescopes} catalog\footnote{http://archive.stsci.edu}). The star is approximately 5.05$\,$pc away from the Sun \citep{GAIADR3}. EV Lac has rotation period of 4.359~days \citep{Paudel_2021}. The surface of EV Lac hosts extensive starspots capable of covering a significant portion of the stellar surface \citep{Jeffers_2022, Ikuta_2023}, and it experiences superflares with energies significantly exceeding those of the strongest solar flares \citep{Ikuta_2023}.

\subsection{Shumen observations}
The data from Shumen Observatory, Bulgaria were collected using a 40~cm reflector Meade LX200ACF equipped with an FLI PL09000 CCD camera \citep{Kjurkchieva_2020}. This configuration provides a $30' \times 30'$ field of view in the $g'$, $r'$, and $i'$ bandpasses of the Sloan photometric system. Due to the absence of significant signal increases in the $i'$ bandpasses during the flare, we limited our analysis to the $g'$ and $r'$ bandpasses.

\begin{figure}[ht!]
    \centering
    \includegraphics[width=0.97\columnwidth]{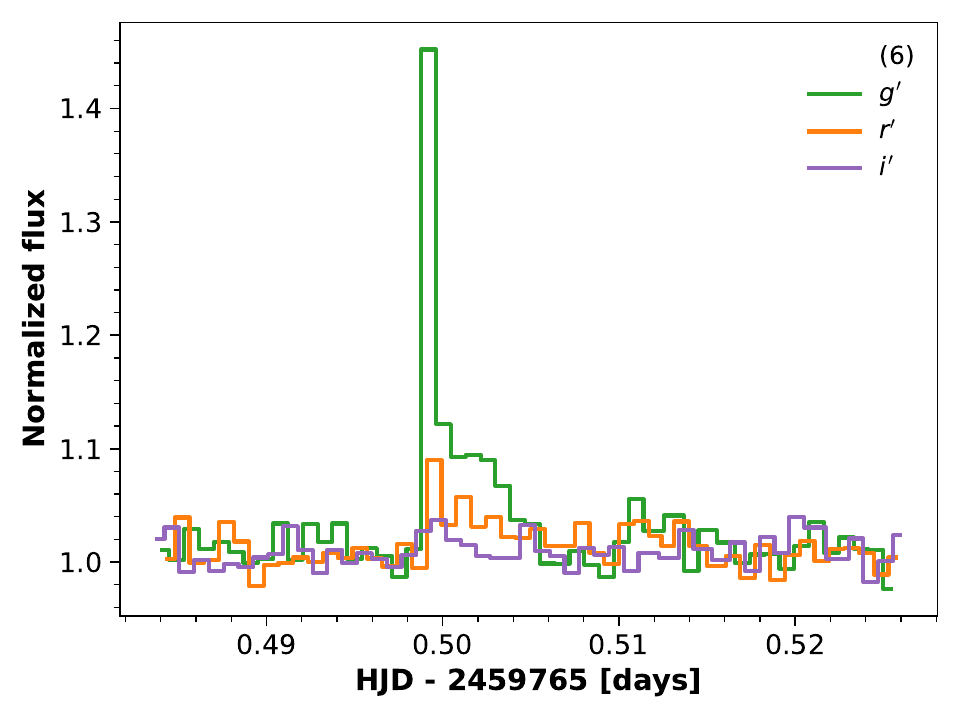}
    \caption{Normalized light curve of the flare No. 6 of V374~Peg in three bandpasses $g'$, $r'$, and the $i'$ (labeled). Observations were made in Shumen Observatory.}
    \label{fig:szumenv}
\end{figure}

\begin{table*}[bp]
\caption[]{Informations about observations of the flares analyzed in this paper.}\label{tab:obsbialshu}
\centering
\small
\begin{tabular*}{\hsize}{@{\hspace{0.2cm}}@{\extracolsep{\fill}}cccc ccc}
\hline\hline\noalign{\smallskip}
Flare \#  & Observational night & Observatory & Star & Used bandpasses & Bandpasses cadences & TESS sector \\
 & [YYYY-MM-DD/DD] & & & & [s] & \\
\noalign{\smallskip}\hline\noalign{\smallskip}
1 & 2022-10-06/07 & Bia\l{}k\'ow & EV Lac & B, {\it TESS} & 70, 20 &   57\\
2 & 2022-10-06/07 & Bia\l{}k\'ow & EV Lac & B, {\it TESS} & 70, 20  &   57\\
3 & 2022-10-06/07 & Bia\l{}k\'ow & EV Lac & B, {\it TESS} & 70, 20 &   57\\
4 & 2022-10-21/22 & Bia\l{}k\'ow & EV Lac & B, {\it TESS} & 70, 20 &   57\\
5 & 2022-10-21/22 & Bia\l{}k\'ow & EV Lac & B, {\it TESS} & 70, 20 &   57\\
6 & 2022-07-04/05 & Shumen & V374 Peg & $g'$, $r'$ & 70, 70  & --- \\
7 & 2022-08-02/03 & Shumen & GJ 1243 & $g'$, $r'$, {\it TESS} & 70, 70, 20 & 54\\
\noalign{\smallskip}\hline\noalign{\smallskip}
\end{tabular*}
\end{table*}

On the night of July 4/5, 2022, a flare on V374 Pegasi was observed (flare No. 6, see Fig. \ref{fig:szumenv}). V374 Peg is a fully convective M3.5Ve dwarf star \citep{Morin_2008b} situated at a distance of 9.1~pc. It has a mass of $0.29 \pm 0.02$~M$_\odot$, a radius of $0.31 \pm 0.01$~R$_\odot$, and an effective temperature of $3\,200 \pm 200$~K (MAST catalog). The estimated rotation period for V374 Peg is $0.4457572 \pm 0.0000002$ days \citep{Bicz_2022}. We observed the star with exposure times equal to 25$\,$s in $g'$ and 20$\,$s in $r'$. Only this star had observations in $i'$ bandpass. The surface of V374~Peg can be spotted over five percent \citep{Bicz_2022}, the spots can be stable on a one-year timescale \citep{Morin_2008}, and energies of the flares on this star reach over $10^{33}$~erg \citep{Bicz_2022}.

\begin{figure}[ht!]
    \centering
    \includegraphics[width=\columnwidth]{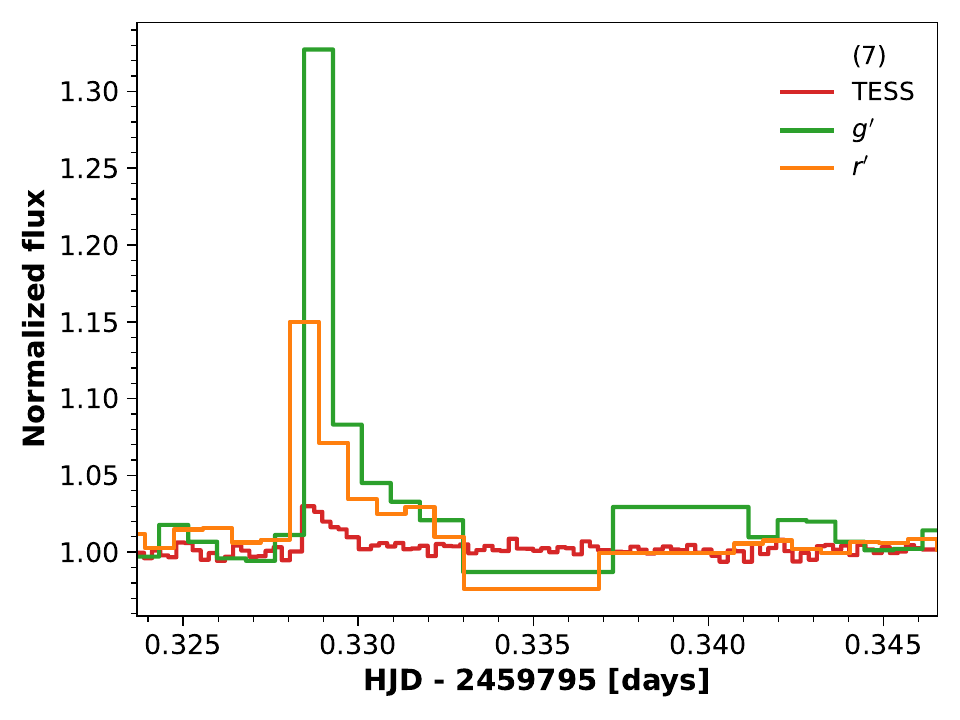}
    \caption{Normalized light curve of the flare No. 7 of GJ~1243, observed in two bandpasses, $g'$ and $r'$ (labeled), at Shumen Observatory, along with the 20$\,$s normalized light curve from {\it TESS} (labeled).}
    \label{fig:szumengj}
\end{figure}

Additionally, on the night of August 2/3, 2022, a flare on the star GJ 1243 was observed at Shumen Observatory (flare No. 7, see Fig. \ref{fig:szumengj}). GJ 1243 is a fully convective M4.0V dwarf star \citep{Lepine_2013} located at a distance of 11.95~pc. The star has a mass of $0.24 \pm 0.02$~M$_\odot$, a radius of $0.27\pm 0.01$~R$_\odot$, and an effective temperature of $3\,200 \pm 200$~K (MAST catalog). Its rotation period is $0.59260 \pm 0.00021$ days \citep{Davenport_2015}. The star had exposure times equal to 20$\,$s in $g'$ and 15$\,$s in $r'$. GJ 1243 is a subject of many studies due to its enormous flaring activity \citep{Davenport_2020, Bicz_2022} and the presence of very long-living spots on its surface \citep{Davenport_2014, Bicz_2022}. The spotedness on this star varies around a few percent and flares have energies that exceed the Carrington Event \citep{Davenport_2014, Bicz_2022}. 

\subsection{{\it TESS} observations}
{\it TESS} is a space telescope launched in April 2018 and placed in a highly elliptical orbit with a period of 13.7~days. The primary objective of the mission is to conduct continuous observations of a substantial portion of the celestial sphere, which is divided into sectors. The satellite monitors target stars with a 2-minute cadence (short cadence) and a 20-second cadence (fast cadence) over a monitoring period of approximately 27 days in each sector. Light curves can also be obtained from the Full Frame Images.

The five flares observed on EV Lac at Bia\l{}k\'ow Observatory were also detected by {\it TESS} during Sector 57. Additionally, the flare on GJ 1243 observed at Shumen Observatory was captured by {\it TESS} in Sector 54. For these six flares, both 20-second and 2-minute cadence data from {\it TESS} observations were available. The compheresive comparision of all conducted observations can be seen in Table \ref{tab:obsbialshu}.

\subsection{Flare data from \cite{Howard_2020}}

To increase our sample of flares we used the continuum temperature evolution data from eight flares analyzed by \cite{Howard_2020} using observations from {\it TESS} and Evryscope \citep{Law_2016, Ratzloff_2019}. We selected the data based on acceptable measurement errors to ensure reliable results. These flares are highlighted in the two bottom panels of Fig. 4 and are labeled (a), (b), (d), (f), (g), and (h) in Fig. 5 of the mentioned paper. Table \ref{tab:howard} shows in detail on which star each flare occurred, the time of occurrence, and the peak temperature (average temperature within the full width at half maximum as defined in \cite{Howard_2020}). Additionally, the first column of Table \ref{tab:howard} shows the flare's number in this paper.
\begin{table}[ht!]
\small
\caption{Information about the flares from \cite{Howard_2020} used in the present paper. The date is expressed in a format YY-MM-DD hh:mm.}
\label{tab:howard}      
\begin{tabular}{c c c c c}          
\hline\hline                        
Flare & TIC & Obs. time & Peak $T_{\rm eff}$  & SpT.\\    
 \# & & (UTC) & [K] & \\
\hline                                   
     8 & 220433364 & 18-11-03 02:35 & $9\,400 \pm 300$  & M4Ve$^{1}$\\ 
     9 & 140045538 & 18-08-14 01:52 & $7\,500 \pm 500$ & M2Ve$^{2}$\\  
    10 & 294750180 & 18-10-20 05:36 & $34\,000 \pm 2\,300$ & M1Ve$^{2}$\\
    11 & 229807000 & 18-08-19 07:46 & $15\,500 \pm 400$ & M2.5Ve$^{3}$\\
    12 & 294750180 & 18-08-20 07:23 & $14\,400 \pm 400$ & M1Ve$^{2}$\\
    13 & 441398770 & 18-08-12 01:57 & $14\,500 \pm 3\,000$ & M4.5Ve$^{4}$\\
    14 & 388857263 & 19-06-03 04:37 & $8\,100 \pm 400$ & M5.5Ve$^{5}$\\
    15 & 201919099 & 18-10-05 08:23 & $9\,700 \pm 300$ & K5Ve$^{2}$\\
\hline                                             
\end{tabular}
$^{1}$\cite{Henry_2002} \\
$^{2}$\cite{refId0}\\
$^{3}$\cite{Riedel_2017}\\
$^{4}$\cite{Joy_1974}\\
$^{5}$\cite{Bessell_1991}
\end{table}

\subsection{Cadence of observations}

Observations in multiple filters were conducted alternately, meaning they are not simultaneous. As a result, it was necessary to align both datasets on a common timeline. To achieve this, a linear interpolation of observation epochs, brightness values, and their uncertainties was performed for data from both observatories (Shumen and Bia\l{}k\'ow) to match the observation times of TESS. Similarly, the Evryscope observations from \cite{Howard_2020} were interpolated onto the TESS timeline.
As an alternative approach we could compare the closest data points from Shumen or Białków observations with the nearest TESS observations, assuming that the physical conditions of events separated by less than a minute remains synchronous. However, this method would have increased the uncertainty in temperature estimations and introduced distortions, either overestimating or underestimating the values depending on the flare phase.
We chose linear interpolation as it eliminates temporal differences between observations from different observatories. Moreover, linear interpolation does not significantly alter the profiles of most observed events, which often last for a considerable duration. Near the peak of the flare's emission, this interpolation method may slightly underestimate the maximum temperature. However, for time differences of only a few seconds, the underestimation of temperature remains negligible.

\section{Methods}\label{sec:methods}
	
	\subsection{Flare temperature and area estimation}\label{subsec:ft}
	The method we used closely follows the approach outlined by \cite{Hawley_2003}. We define the temperature of a flare ($T_{\rm flare}$) as the effective temperature deduced from the flare's spectral characteristics. The {\it TESS} bandpass lies in the tail of the Planck curve, resulting in minimal energy excess within the {\it TESS} bandpass for high temperatures (temperatures higher than 10$\,$000$\,$K). 
However, other bandpasses used in observations fall in the bluer region of the Planck curve, leading to higher flare amplitudes at elevated temperatures. 
Some of the spectroscopic observations of flares show a non-negligible Balmer jump, but focusing the energy estimates on optical continuum which provides a dominant contribution, we consider the black-body temperature obtained from that continuum. Both spectral observations as well as RHD simulations show, for very strong electron-beam fluxes, that the Paschen optical continuum can be well fitted by the Planck function, while the Balmer continuum is present apparently due to non-homogeneous temperature and density structure of the chromospheric condensation (see \cite{Kowalski2024}). Note that at lower densities of the chromospheric condensations the Balmer jump appears naturally even in the case of isothermal slabs (e.g. \cite{Heinzel_2024}). Moreover, the ultraviolet tail of the $B$ filter from the Białków Observatory that we used exhibits a minimal response to the Balmer jump, approximately 0.4\%.
The temperature estimation process is as follows:
	\begin{enumerate}
	\item We use synthetic spectra from the PHOENIX library \citep{phoenix} to obtain the radiation spectrum for a star with the specified effective temperature, metallicity, and surface gravity over the wavelength range encompassing the entire filter in which the flare was observed.
	\item We multiply the spectrum by the response functions for the given filters and integrate over all wavelengths to determine the star's flux in each bandpass and the energy emitted by the flare at each instant.
	\item By estimating the energy surpluses in multiple bandpasses, we calculate the flare's area and temperature at each measurement point by solving the equations $L_{\rm flare}~=~A_{\rm flare} B_\lambda(T_{\rm flare})$ for the set of bandpasses where $A_{\rm flare}$ is the area of the flare. To solve the equation we used the assumption that the flare's area do not vary much in different bandpasses thus we may solve the
system of equations for $T_{\rm flare}$ and $A_{\rm flare}$ separately \citep{Castellanos_2020, Howard_2020}.
	\end{enumerate}

Using the temporal temperature evolution data from \cite{Howard_2020} as a basis, we analyzed the changes in the flare area in time using the light curve observed by the {\it TESS} satellite. To perform calculations we assume that the spectrum of white-light flares can be modeled as black-body radiation, which is consistent with optically thick Pashen continuum \citep{Heinzel_2024}. 
The combined luminosity ($L$) from both the flare and the remaining stellar surface can be expressed as:
	\begin{equation} 
	    L = \left(A_{\rm star} - A_{\rm flare}\right)F_{\rm star} + A_{\rm flare} 
	         F_{\rm flare}{\rm ,} 	
	\end{equation}
	 	where $F_{\rm flare}$, $A_{\rm flare}$, $F_{\rm star}$, and $A_{\rm star}$ represent the fluxes and areas of the stellar disk and the flare, respectively. The flux of the star is described by:
	\begin{equation} 
	 	L_{\rm star} = A_{\rm star} F_{\rm star}.
	\end{equation}
	The normalized light curve of the star ($C$) is then given by:
	\begin{equation} 
	C = \frac{L}{L_{\rm star}} = \frac{\left(A_{\rm star} - A_{\rm flare}\right)F_{\rm star} + A_{\rm flare} F_{\rm flare}}{A_{\rm star} F_{\rm star}}.
	 \end{equation}
	This formula can be rewritten to express the area of the flaring region as:
	\begin{equation} 
		A_{\rm flare} = \frac{(C-1)A_{\rm star}F_{\rm star}}{F_{\rm flare} - F_{\rm star}}. 
	\end{equation}
	The above relation can naturally be rewritten as:
	\begin{equation}\label{eq:aflare}
		A_{\rm flare} = \frac{(C-1) \pi R_{\rm star}^2 \int\limits_{\lambda_1}^{\lambda_2} S_{\lambda} B_\lambda(T_{\rm star}) d\lambda}{\int\limits_{\lambda_1}^{\lambda_2} S_{\lambda} B_\lambda(T_{\rm flare}) d\lambda - \int\limits_{\lambda_1}^{\lambda_2} S_{\lambda} B_\lambda(T_{\rm star}) d\lambda}, 
	\end{equation}
where $\lambda$ is the wavelength, $B_\lambda(T)$ is the Planck function, $T_{\rm star}$ and $R_{\rm star}$ are the star's effective temperature and radius, respectively, and $S_{\lambda}$ is the instrument's response function. This equation makes possible for us to estimate how the flare area evolves throughout the event, provided the temporal temperature evolution and photometric data.

\subsection{Observational estimation of flare peak temperatures}\label{sec:aaa}

Available flare observations allow us to estimate the flare temperature only at the observational points relative to a chosen filter. As a result, we often lack an exact measurement of the peak signal, since the maximum phase is short-lived. By fitting a theoretical flare profile to observations, we can infer where this maximum should occur.

We fitted the Wroc\l{}aw Flare Profile \citep{Gryciuk_2017, Bicz_2022, Pietras_2022} to the photometric data using an evolutionary algorithm code\footnote{https://github.com/KBicz/Wroclaw-Flare-Profile-Fitter} \citep{Bicz_20242} to model the temperature evolution of the analyzed flares. For all flares except No. 6, we utilized the {\it TESS} light curve, selected for its high data cadence and low noise (this flare was not observed by {\it TESS}). For flare No. 6, we chose the $g'$ light curve (represented by the green curve in Fig. \ref{fig:szumenv}). Each flare was subjected to 10$\,$000 fits to achieve the best possible match. By considering possible peak temperatures—from the star's effective temperature up to 60$\,$000$\,$K—along with the estimated flare surface from multi-color photometry we can estimate the theoretical temperature evolution of the flare using fitted profile by solving the rewritten Eq.~\ref{eq:aflare}:
	\begin{equation}\label{eq:integral}
        \int\limits_{\lambda_1}^{\lambda_2} S_{\lambda} B_\lambda(T_{\rm flare}) d\lambda = \left(\frac{(C-1) \, \pi R_{\rm star}^2}{A_{\rm flare}} + 1\right)\int\limits_{\lambda_1}^{\lambda_2} S_{\lambda} B_\lambda(T_{\rm star}) d\lambda ,
    \end{equation}
Using this grid of solutions for a given range of peak tempratures we can find the best fit to the temperature evolution data (estimated as in Sect. \ref{subsec:ft}). This allows us to deduce the flare's peak temperature. We used the normalized $\chi^2$ statistics to evaluate the goodness of fit.

\begin{figure*}[ht!]
    \centering
    \includegraphics[width=0.95\textwidth]{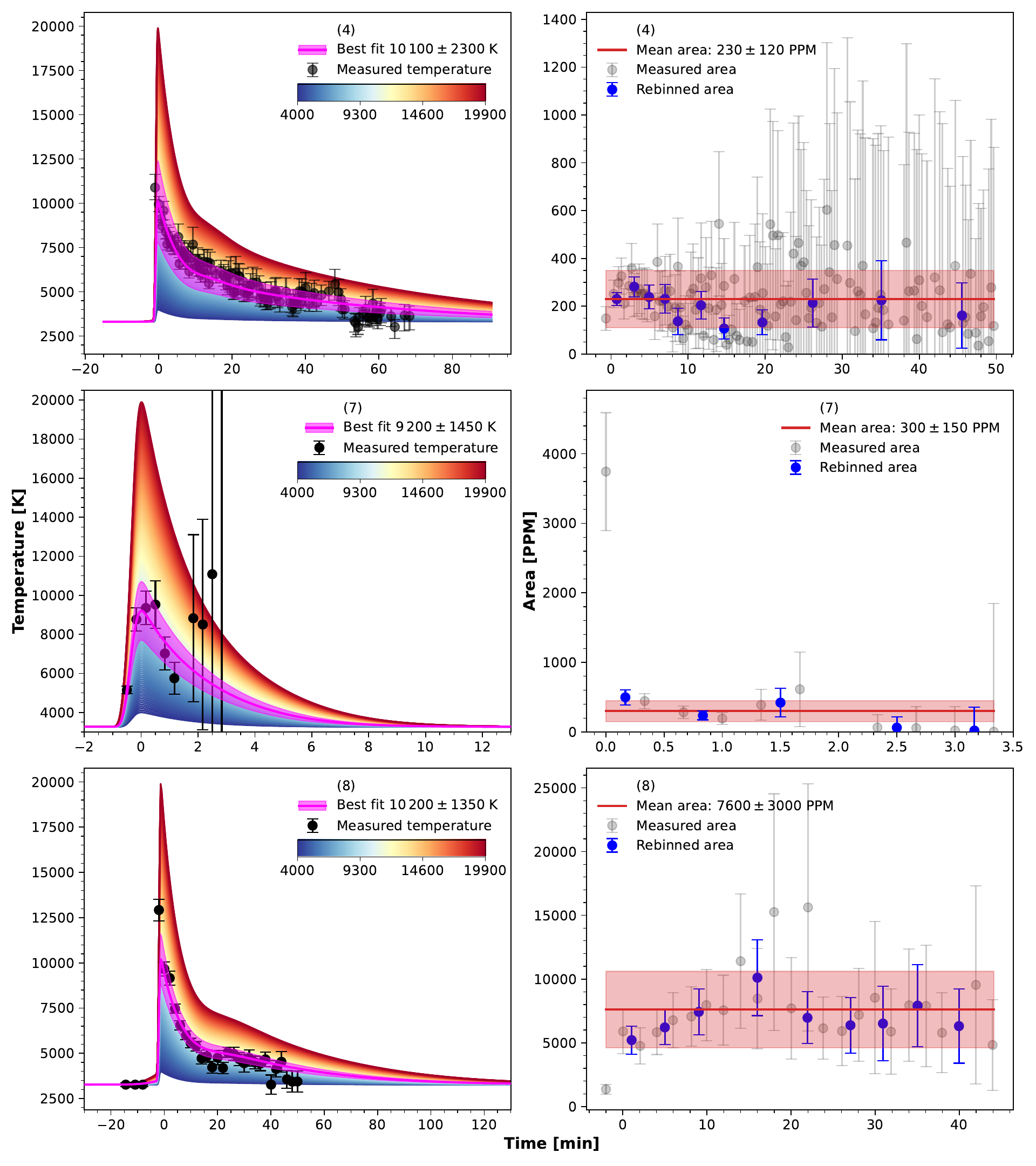}
    \caption{Measured temperature evolution (left), and the evolution of the flare areas (right) for the flares No. 4 (top panels), No. 7 (middle panels), and No. 8 (bottom panels). In the left panels, the black dots represent the temperature evolution determined from the multi-color observations. The color gradient area depicts the black-body temperature evolution throughout the event for various peak temperatures of the flares. The magenta line and the shaded interval indicate the best fit and its associated uncertainty range. In the right panels, the gray dots show the flare area evolution over the entire event, with the red line representing the mean area and the shaded red interval indicating its uncertainty. Blue points with error bars show the rebinned area measurements throughout the event.  The values on abscissa show time in minutes from the peak of the flare. The number in the upper part of each panel represents the number of the flare in this paper (Table \ref{tab:theoobs}).}
    \label{fig:restemparea}
\end{figure*} 

\subsection{Flare energy estimation}\label{subsec:fenergy}

The bolometric energy of each flare ($E_{\rm flare}$) can be calculated using the temporal temperature evolution of the flare ($T_{\rm flare}$) with the corresponding flaring area ($A_{\rm flare}$). This calculation involves summing the energy emitted at different time intervals ($\Delta t$) across the entire electromagnetic spectrum, assuming that the flare's spectrum can be modeled as black-body radiation of a given temperature $T_{\rm flare}$. The expression for $E_{\rm flare}$ (Method I) is given by:
\begin{equation}\label{eq:eq7}
	E_{\rm flare} = \sum\limits_{i \, = \, 0}^{n-1} \left( A_{{\rm flare},\, i}\,\Delta t_i\, \pi \int\limits_0^{+\infty} B_\lambda(T_{{\rm flare,} \, i}) \, d\lambda \right),
\end{equation} 
where $B_\lambda$ is the Planck function and $n$ is the total number of measurement points captured during the flare. To further compare the results of Method I, we used the approach outlined by \cite{Shibayama_2013} (referred to as Method II), which assumes a constant flare temperature of $10\,000\,$K while allowing the flare area to vary. Method II also assumes that the flaring region is substantially smaller than the total stellar surface.

\section{Results}\label{sec:results}

\subsection{Evolution of observational areas and temperatures of flares}\label{subsec:evol}
We found that the estimated peak black-body temperatures of our newly observed flares ranged from $5\,700\,$K to $17\,500\,$K (Table \ref{tab:theoobs}). To validate our flare peak temperature estimation method, we applied it to the eight flares analyzed from \cite{Howard_2020}. The peak flare temperatures for the flares from \cite{Howard_2020} estimated by us ranged from $7\,000$ to $31\,500\,$K, and all our peak temperature estimates for these flares are consistent with their results within the margin of error.
Among our newly observed flares, two had peak temperatures below $8\,000\,$K, four fell within the $8\,000\,$K$\,$–$\,14\,000\,$K range (the upper limit of the typical range quoted in \cite{Kowalski_2013}), and one exceeded $14\,000\,$K.
The bolometric energies of our seven flares, estimated using Method I, ranged from \flaree{2.0}{1.3}{31}$\,$erg to \flaree{1.9}{0.9}{32}$\,$erg, while for the flares from \cite{Howard_2020}, they ranged from \flaree{3.3}{1.7}{31}$\,$erg to \flaree{1.2}{0.5}{32}$\,$erg. 
All seven flares observed in our study had energies below the superflare threshold ($10^{33}\,$erg). In contrast, all flares that we used in this paper from \cite{Howard_2020} exceeded this threshold, except for flares 9 and 14. 
The bolometric energies derived using Method II ranged from \flaree{1.9}{0.6}{31}$\,$erg to \flaree{1.9}{0.5}{35}$\,$erg for our seven observed flares and from \flaree{3.3}{1.7}{31}$\,$erg to \flaree{1.2}{0.5}{35}$\,$erg for the flares analyzed by \cite{Howard_2020}.

\begin{figure}[ht!]
    \centering
    \includegraphics[width=\columnwidth]{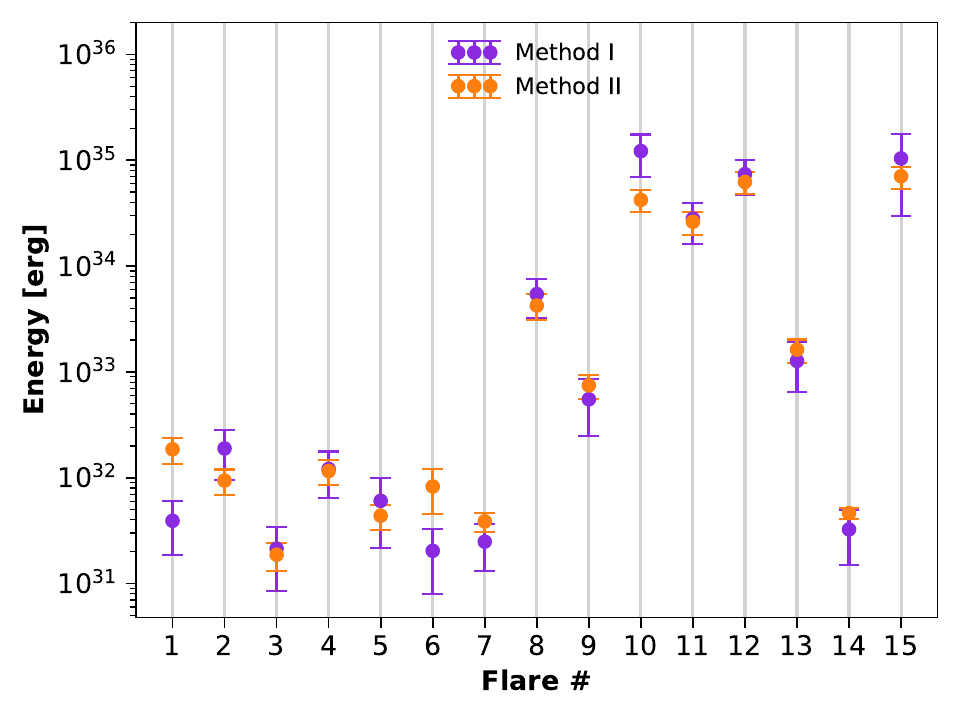}
    \caption{Comparison of the flare energies estimated using Method I (violet points), and Method II (orange points).}
    \label{fig:comparision}
\end{figure}

For the flares No. 1, 6, and 10, Method II produced significantly different energies, with differences ranging from the energy 0.65 times lower for flare No. 10 to the energy approximately 4 times higher for flare No. 1.
The average flare areas ranged from $50\,\pm\,30\,$ppm\footnote{Parts per million of a stellar disk} for flare No. 6 to 300$\,\pm\,$150$\,$ppm for flare No. 7 among our seven newly observed flares, and from 380$\,\pm\,$200$\,$ppm for flare No. 14 to $7600\,\pm\,3000\,$ppm for flare No. 8 in the data from \cite{Howard_2020}. Only one flare had an area smaller than $100\,$ppm; nine flares had areas between $100\,$ppm and $1000\,$ppm, and five flares had areas exceeding $1000\,$ppm. To improve the clarity of the flare area results and reduce the influence of the very large observational uncertainties an adaptive rebinning was performed on the data. Data points were grouped into different-sized bins depending on the associated uncertainties: in regions which had small uncertainties, few bins were grouped together, while larger bins were used in areas of higher uncertainty. Again, data points where the relative uncertainty became larger than the value itself had their bin size further increased to receive uncertainty lower than the rebinned value (see Fig. \ref{fig:restemparea}). During the rebinning process the data uncertainties were used as a weights. This rebinning allowed us to estimate that the flare area varies, on average, by approximately 50\% of the mean flare area value. For almost all of the analyzed flares we remark the lack of significant changes of the estimated flare area. Some variations are noted for the rise phases of the flares 1, 8 and 11. When the flares pass to the gradual phase, these changes vanish. It is worth to note, that although it is physically plausible that a flare develops during the rise phase, this phase of the evolution is hard to control with the time cadence achieved in our observations. We performed statistical tests to evaluate the validity of the assumption that these areas remain constant. Column 4 in the Table \ref{tab:theoobs}  lists the value of possible changes of the flare area estimated from the linear fit to the rebinned data relative to their weighted average. This value usually does not exceed the level of 30\%. Only for the flares 5 and 7 it reaches the value of over 50\%. We also performed F-test to compare the residual variances for the two assumptions: first that the area is constant and second that it shows an important trend. Column 5 lists the probability of the assumption that there are no significant changes in the estimated area. As one can see this probability is high for almost all analyzed flares. The only outstanding event is flare No. 7, but even here this probability is not low enough to statistically reject the assumption of the constant area model.
\begin{table*}[ht!]
\caption[]{Peak temperatures and mean areas of the analyzed flares, along with the p-value of the F-statistic used to assess whether the slope of the regression is statistically significant, determining if the flare area remains constant over time. The relative area change below flare No. 7 separates our observed flares from data from \cite{Howard_2020}.}\label{tab:theoobs}
\centering
\small
\begin{tabular*}{\hsize}{@{\hspace{0.2cm}}@{\extracolsep{\fill}}ccccccc}
\hline\hline\noalign{\smallskip}
Flare \#  & $T_{\rm flare}$ & $A_{\rm flare}$ & Relative area change & p-value & Method I & Method II \\
 & [K] & [ppm] & [\%] &  & [erg] & [erg] \\
\noalign{\smallskip}\hline\noalign{\smallskip}
1 &  $5\,700 \pm 450$ & $210 \pm 120$ & 17 & 0.15 & \flaree{3.9}{2.1}{31} & \flaree{1.9}{0.5}{32} \\
2 &  $\,13\,900 \pm 1\,900$ & $370 \pm 180$ & 33 & 0.13 & \flaree{1.9}{0.9}{32} & \flaree{9.4}{2.6}{31} \\
3 &  $\,\,\,\,6\,400 \pm 1\,800$ & $150 \pm 90\,\,\,$ & 43 & 0.49 & \flaree{2.1}{1.3}{31} & \flaree{1.9}{0.6}{31} \\
4 &  $\,10\,100\pm 2\,300$  & $230 \pm 120$ & 10 & 0.64 &  \flaree{1.2}{0.6}{32} & \flaree{1.2}{0.3}{32} \\
5 &  $\,11\,100 \pm 3\,150$ & $220 \pm 140$ & 61 & 0.24 & \flaree{6.1}{3.9}{31} & \flaree{4.4}{1.2}{31} \\
6 &  $\,17\,500 \pm 10\,000$ & $50 \pm 30$ & 33 & 0.57 & \flaree{2.0}{1.3}{31} & \flaree{8.3}{3.8}{31} \\
7 & $\,\,\,\,9\,200 \pm 1\,450$ & $300 \pm 150$ & 51 & 0.45 & \flaree{2.5}{1.2}{31} & \flaree{3.9}{0.8}{31} \\
\hline\noalign{\smallskip}
8 & $\,10\,200 \pm 1\,350$ & $7\,600 \pm 3\,000$ & 13 & 0.38 & \flaree{5.4}{2.2}{33} & \flaree{4.2}{1.2}{33} \\
9 & $\,\,\,\,7\,000 \pm 3\,100$ & $440 \pm 240$ & 18 & 0.63 & \flaree{5.5}{3.0}{32} & \flaree{7.5}{1.9}{32} \\
10 & $\,\,\,\,31\,500 \pm 10\,050$ & $3\,200 \pm 1\,200$ & 14 & 0.46 & \flaree{1.2}{0.5}{35} & \flaree{4.2}{1.0}{34} \\
11 & $\,18\,200 \pm 2\,850$ & $2\,900 \pm 1\,200$ & 41 & 0.07 & \flaree{2.8}{1.2}{34} & \flaree{2.6}{0.6}{34} \\
12 & $\,15\,400 \pm 3\,550$ & $4\,700 \pm 1\,400$ & 19 & 0.11 & \flaree{7.4}{2.7}{34} & \flaree{6.2}{1.5}{34} \\
13 & $12\,600  \pm 7\,050$ & $410 \pm 160$ & 27 & 0.14 & \flaree{1.3}{0.6}{33} & \flaree{1.6}{0.4}{33} \\
14 & $\,\,\,7\,500 \pm 2\,350$ & $380 \pm 200$ & 39 & 0.21 & \flaree{3.3}{1.7}{31} & \flaree{4.6}{0.6}{31} \\
15 & $12\,500 \pm 2\,650$ & $2\,200 \pm 1\,500$ & 10 & 0.51 & \flaree{1.0}{0.7}{35} & \flaree{7.0}{1.7}{34} \\
\noalign{\smallskip}\hline\hline\noalign{\smallskip}
\end{tabular*}
\end{table*}
The temperature and area evolution for selected flares are illustrated in Fig. \ref{fig:restemparea}.
Table \ref{tab:theoobs} presents the detailed values of bolometric energies (see also Fig. \ref{fig:comparision}), estimated areas, and peak flare temperatures, along with their uncertainties. Notably, only two flares (1 and 7) required a single-component Wroc\l{}aw flare profile for a good fit, while the remaining flares required a two-component Wroc\l{}aw flare profile which is consistent with the previously obtained results (see \cite{Bicz_2022} and \cite{Pietras_2022} for more details).

\subsection{Semi-empirical grid for estimating the peak flare temperatures of main-sequence star}

\begin{figure}[ht!]
    \centering
    \includegraphics[width=\columnwidth]{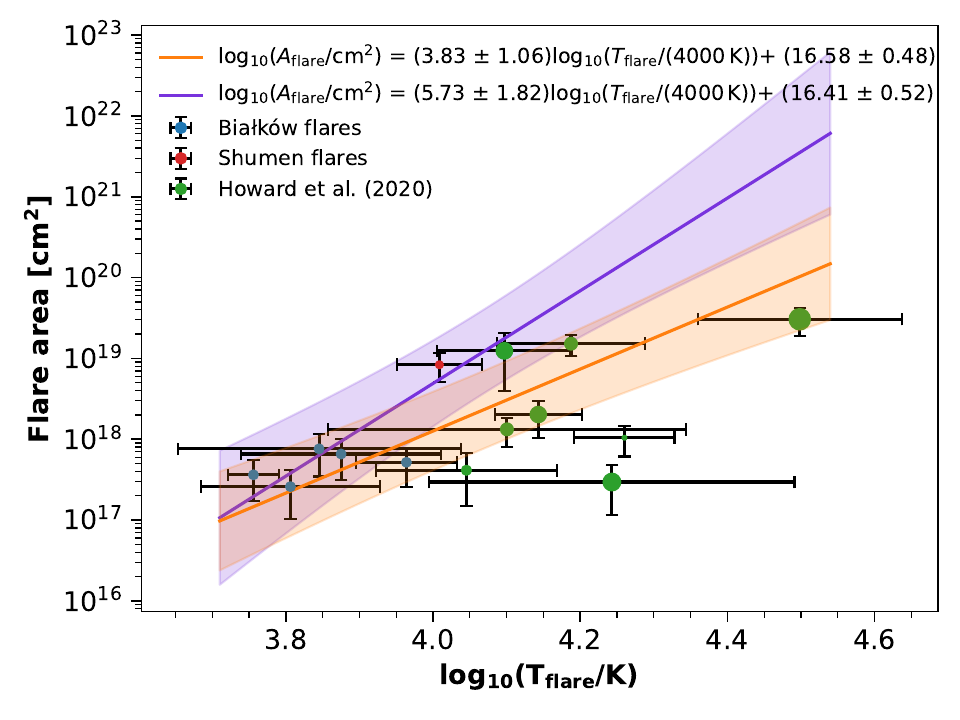}
    \caption{Relationship between flare area and flare peak temperature $T_{\rm flare}$. Flares observed at the Bia\l{}k\'ow Observatory are shown in blue, while data from \cite{Howard_2020} are represented in green. The flare from the Shumen Observatory is marked in red (flare No. 7). The size of the point indicates the radius of the star on which the flare occurred. The purple and the orange curves, along with their shaded regions, illustrate the power-law fit to our observational data and our observational data combined with the data from \cite{Howard_2020} respectively.}
    \label{fig:corelat}
\end{figure}

\begin{figure}[ht!]
    \centering
    \includegraphics[width=\columnwidth]{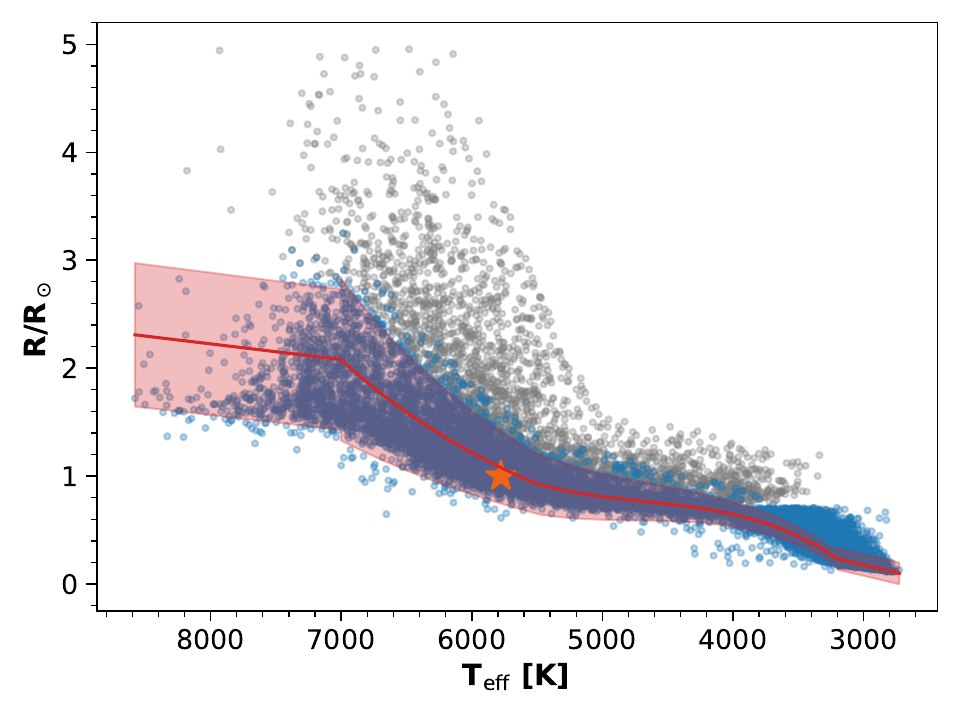}
    \caption{Relation between stellar radius and effective temperature for the 16$\,$000 most active main-sequence stars from \cite{Pietras_2022} (blue dots). The red line presents the fit described by Eqs. \ref{eq:rstar1} - \ref{eq:rstar4}, the red interval represents the fit error and the orange star corresponds to the parameters of the Sun. The gray points represent possible falsely detected main-sequence stars in the MAST catalogue (2$\,$000 stars).}
    \label{fig:hr}
\end{figure}

Most of the stellar flares are observed only in one bandpass. This includes huge database of flares recorded by {\it Kepler} and {\it TESS} missions. For such events, it is not possible to apply the analysis outlined in this paper. To make it possible we must estimate the flare temperature at its peak. In this study, we aim to develop a method for making reasonable predictions of the peak temperatures of the flares basing on the stars spectral classification and the observed flares magnitudes.
The observed flux from a flare is influenced by both the size of the flaring region and the time-dependent evolution of its temperature (see Sect. \ref{subsec:ft}, point 3). Typically, estimating the flare's peak temperature requires knowing the area of the flare at its peak and the corresponding flux increase. However, since the peak temperature of a flare shows a strong positive correlation with its area (correlation probability, p = 0.98; see Fig. \ref{fig:corelat}), it may be possible to reconstruct the peak temperature without estimating the flare's area directly.
To achieve this, we analyzed 6 flares from this paper and 8 flares from \cite{Howard_2020}, all of which had available {\it TESS} light curves. By examining the black-body temperature evolution of the flares ($T_{\rm flare}$) and the flare areas ($A_{\rm flare}$), we were able to estimate the luminosity ($L_{\rm flare}$) of each flare in the {\it TESS} bandpass using the following equation:
\begin{equation}\label{eq8}
	L_{\rm flare} = A_{\rm flare}\int\limits_{\lambda_1}^{\lambda_2} B_\lambda (T_{\rm flare}) S_{\rm TESS} \, d\lambda,
\end{equation}
where $B_\lambda$ is the Planck function and $S_{\rm TESS}$ is the response function of {\it TESS} satellite. 
We calculated the luminosities of stars over a temperature range from 2700$\,$K to 4$\,$600$\,$K, with 100$\,$K increments  to estimate what would be the relative luminosities of our flares on different main-sequence flaring stars (what amplitude these flares would have). 
We estimate luminosities only for temperatures up to 4$\,$600$\,$K, as this corresponds to a K4V star \citep{Pecaut_2013}, while the earliest spectral type in our analysis is K5V from \cite{Howard_2020} data. We accomplished this using the equation:
\begin{equation}\label{eq9}
	L_{\rm star} = \pi R_{\rm star}^2\int\limits_{\lambda_1}^{\lambda_2} \mathcal{F}_\lambda S_{\rm TESS} \, d\lambda.
\end{equation}
Where $\mathcal{F}_\lambda$ represents the flux from the theoretical spectrum of a given star from the PHOENIX spectra library. 
Additionally, to estimate how these 14 flares would appear on hypothetical main-sequence flaring stars within the specified temperature range, we assumed that all flares on such stars, regardless of spectral type, follow a standard model of solar flare driven by magnetic reconnection \citep{Shibata_1995, Notsu_2013, Pietras_2022, Pietras_2023, Bicz_2024}.
To use the above equation, we need to know the stellar radius, the logarithm of gravitational acceleration, stellar metallicity, and turbulent velocity to properly synthesize the stellar spectrum and obtain the area of the star. We interpolated the logarithm of gravitational acceleration using the relation between effective temperature and the logarithm of gravitational acceleration as described by \cite{Angelov_1996}. 
We used the relations between stellar radius and effective temperature, described in the four below equations, to determine the radius of the star. We derived first two below equations by refining some of the relations between the radii of main-sequence stars and their effective temperatures for different temperature ranges. This refinement involved improving the fit of the stellar radius as a function of effective temperature for much larger sample of 16$\,$000 most flaring main-sequence stars (using data from \cite{Pietras_2022}) based on their parameters from the MAST catalog (see Fig. \ref{fig:hr}). The 2$\,$000 stars are probably not main-sequence stars and were removed from the further analysis. We estimated the radius-effective temperature relationships for stars with temperatures ranging from 2$\,$700$\,$K to over 4$\,$600$\,$K—extending up to 8$\,$600$\,$K—since the flares on active stars in \cite{Pietras_2022} occurred on main-sequence stars with temperatures up to approximately 8$\,$600$\,$K. Our goal is to enable the extension of our grid to include flaring stars with temperatures beyond our 4$\,$600$\,$K limit.
Below equation presents the refined relation from \cite{Cassini_2019}, for the stars that have effective temperatures in the range $2700$~K $\leq T_{\rm eff} \leq 3200$~K
\begin{equation}\label{eq:rstar1}
R_{\rm star}(T_{\rm eff}) \, [{\rm R_\odot}] = \left(1.644 \left(T_{\text{{eff}}}/5777\right) - 0.677\right) \pm 0.1.
\end{equation}
Similarly, the next equation presents the corrected relation from \cite{Boyajian_2017}, for stars that have effective temperatures in the range $3200$~K $< T_{\rm eff} \leq 5500$~K
\begin{equation}\label{eq:rstar2}
	\begin{aligned}
		R_{\rm star}(T_{\rm eff}) \, [{\rm R_\odot}] = (-10.933 \pm 0.014) \,\, + \\
		+ \,\, (7.188 \pm 0.009) \times 10^{-3} T_{\text{{eff}}} \,\, - \\
			  + \,\, (1.510 \pm 0.002) \times 10^{-6} T_{\text{{eff}}}^2 \,\,  + \\
			  + \,\, (1.076 \pm 0.016) \times 10^{-10} T_{\text{{eff}}}^3.
	\end{aligned}
\end{equation}
\noindent The following equations represent our fitted models for main-sequence stars, each applying to different temperature ranges.
For stars with effective temperatures in the range $5500$~K $< T_{\rm eff} \leq 7000$~K, the stellar radius is given by:
\begin{equation}\label{eq:rstar3}
	\begin{aligned}
		R_{\rm star}(T_{\rm eff}) \, [{\rm R_\odot}] = 10^{(0.000220 \pm 0.000021) \cdot T_{\text{{eff}}} - (1.209 \pm 0.178)} \, - \\
		+ \, (0.080 \pm 0.002).
	\end{aligned}
\end{equation}
For stars with effective temperatures in the range $7000$~K $< T_{\rm eff} \leq 8600$~K, the stellar radius follows the relation:
\begin{equation}\label{eq:rstar4}
R_{\rm star}(T_{\rm eff}) \, [{\rm R_\odot}] = ((14.48 \pm 2.638) \times 10^{-5}) \cdot T_{\text{{eff}}} + (1.067 \pm 0.625).
\end{equation}
Eqs. \ref{eq:rstar3} and \ref{eq:rstar4} are the result of our own fit to the data.
Additionally, to estimate the luminosity of the hypotehtical stars in the mentioned range of temperatures, we used spectra with the assumed solar metallicity and the turbulent velocity of 2 km/s.
For each stellar flux, we incorporated the peak flux from each of the 14 flares and estimated their amplitudes for the respective stars. The relationship between the flare amplitude and the peak temperature of the flare appeared to follow a power-law trend for each star. Consequently, we fitted the following equation to the data:
\begin{equation}
	{\rm log}_{10}(\Delta F_{\rm flare}) = a \cdot {\rm log}_{10}(T_{\rm flare}) + b,
\end{equation}
\begin{figure}[ht!]
    \centering
    \includegraphics[width=\columnwidth]{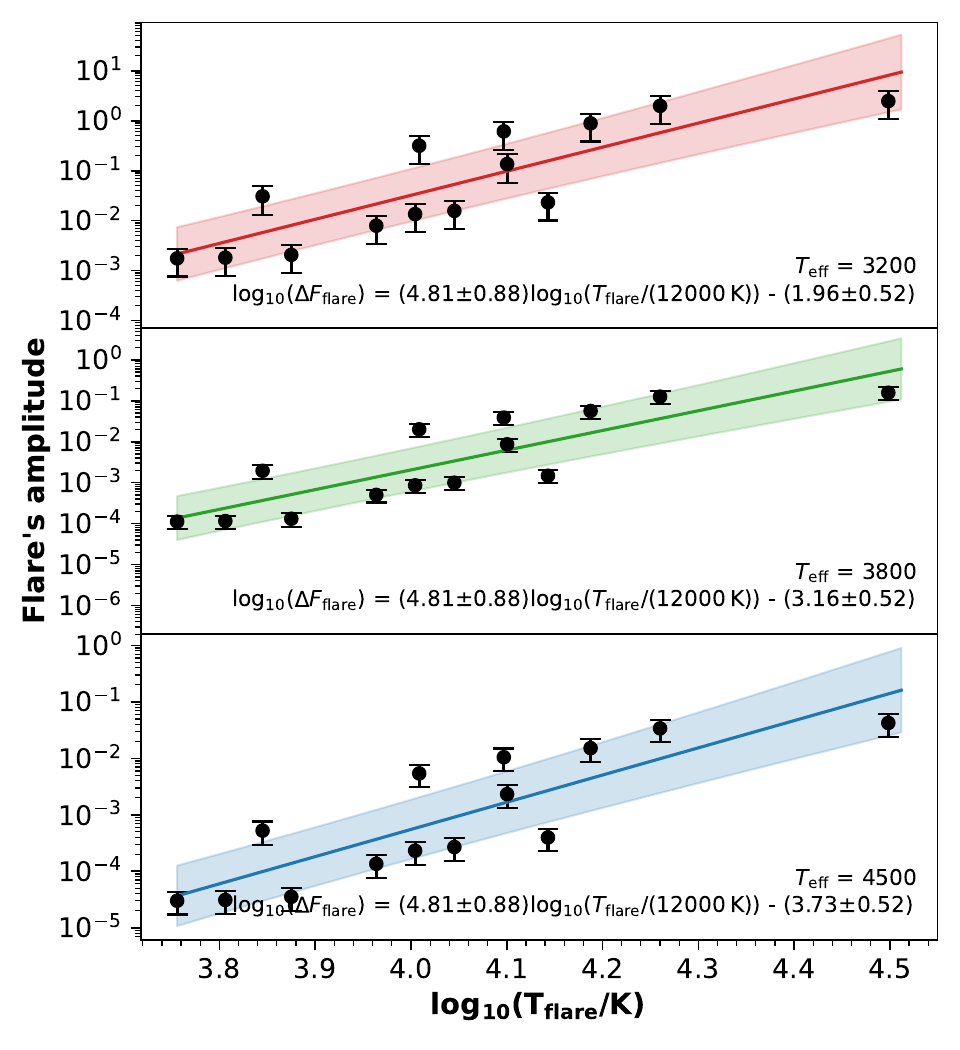}
    \caption{Selected examples of relationships between estimated peak temperatures of the flares $T_{\rm flare}$ and calculated amplitudes of the flares (black dots) for three selected effective temperatures of the stars equal to 3200, 3800, and 4500$\,$K. The red, green, and blue lines and intervals show the power-law fit with fit uncertainty.}
    \label{fig:fity}
\end{figure}
where $\Delta F_{\rm flare}$ is the peak amplitude of the flare. The example fits are illustrated in Fig. \ref{fig:fity}. 
The power law index for the fits is approximately about $4.81 \pm 0.88$. These fits enabled us to construct a grid that describes the relationship between the flare amplitude, stellar effective temperature, and the peak flare temperature for main-sequence flaring stars observed by {\it TESS} (see Fig. \ref{fig:grid}). This grid allows us using interpolation\footnote{https://github.com/KBicz/Flare-Maximal-Temperature-Grid-TESS} to determine the peak flare temperature based on the effective temperature of the star and the amplitude of the flare. The grid provides a reasonably accurate approximation, with  the differences between the grid-estimated and observed values consistently falling within a 30\% margin.

\subsection{Semi-empirical peak temperatures of flares}
Using our semi-empirical grid, we calculated the peak black-body temperatures for the sample of 42$\,$257 flares observed on 268 most active stars from \cite{Pietras_2022}. Only 3\% of these flares had peak temperatures below 7$\,$000$\,$K, 22\% ranged between 7$\,$000$\,$K and 10$\,$000$\,$K, 42\% had peak temperatures between 10$\,$000$\,$K and 14$\,$000$\,$K, 28\% ranged from 14$\,$000$\,$K to 20$\,$000$\,$K, and 5\% exceeded 20$\,$000$\,$K. The lowest estimated peak temperature was 5$\,$700$\,$K, observed in the flare from the M4V star YZ CMi, while the highest temperature reached over 38$\,$000$\,$K in the flare from the M0V star TIC293507177. During the event on YZ CMi, the detected signal from the star increased by 0.2\%, whereas the signal from TIC293507177 increased by approximately 135\%. More than 50\% of the flares had peak black-body temperatures of approximately $11\,100 \pm 2\,400\,$K. The histogram showing this distribution is presented in Fig. \ref{fig:histo}.
\begin{figure}[ht!]
    \centering
    \includegraphics[width=\columnwidth]{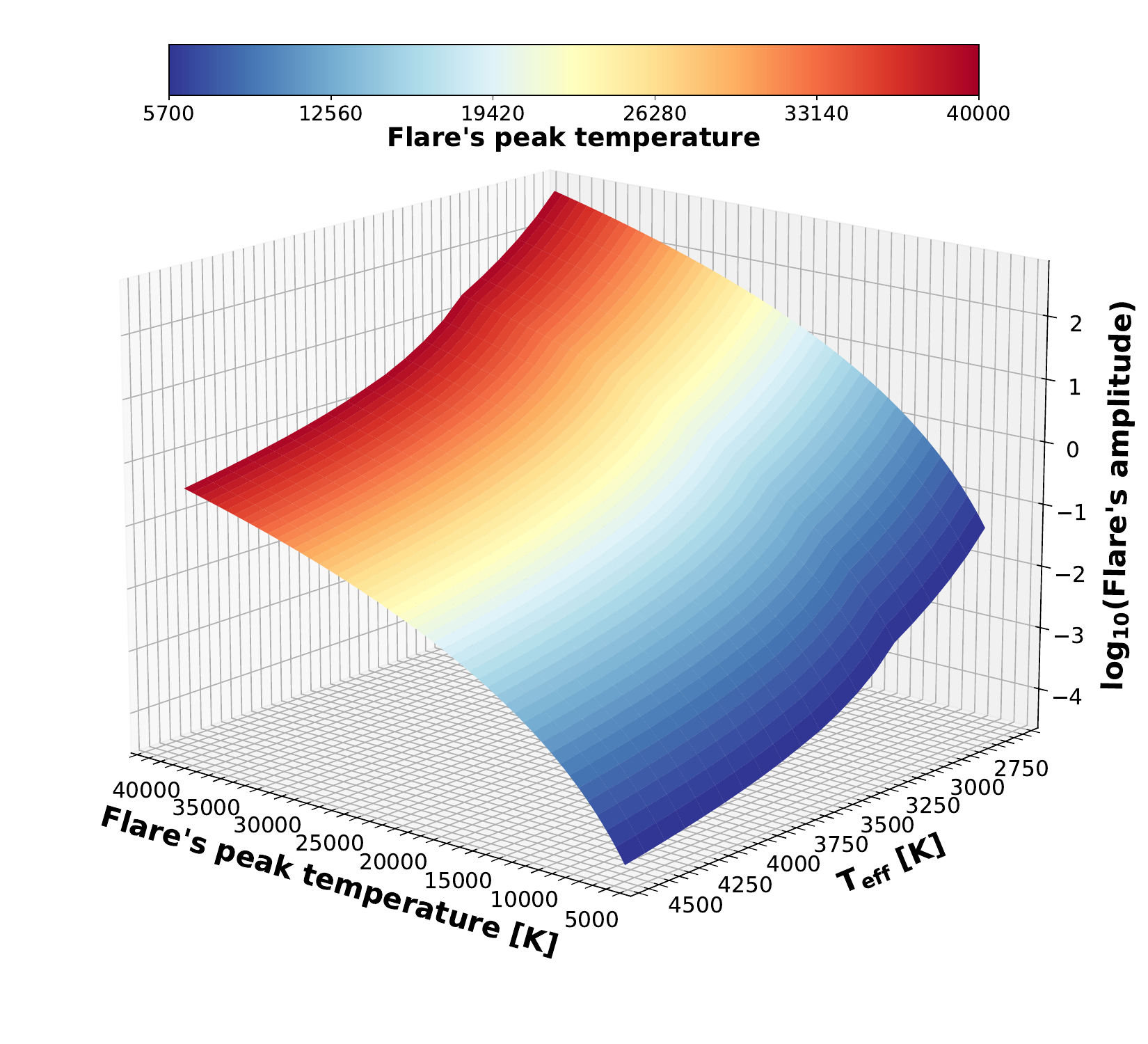}
    \caption{Semi-empirical relationship linking stellar effective temperature, the amplitude of its flares, and the peak temperature of the flares. The color gradient indicates the peak temperature of the flares.}
    \label{fig:grid}
\end{figure}

\section{Summary and Discussion}\label{sec:discussion}

\subsection{Summary of Flare Observations and Parameters}
In this paper, we analyzed 15 stellar flares from 10 different stellar objects using observations from the Bia\l{}k\'ow Observatory, Shumen Observatory,
{\it TESS} satellite, and flare temperature evolution from \cite{Howard_2020}. These stars are main-sequence stars with spectral types ranging from M5.5V to K5V (9 M-dwarfs and 1 K-dwarf). We were able to estimate the evolution of both the flare temperatures and flare areas during the events for the five flares observed at Bia\l{}k\'ow and the two flares observed at Shumen. Additionally, we estimated the area evolution for eight flares using temperature data from \cite{Howard_2020}. The analyzed flares exhibited peak temperatures ranging from $5\,700\pm450\,$K to $31\,500\pm10\,050\,$K, with areas spanning from $50\pm30\,$ppm to $7\,600\pm3\,000\,$ppm. 

For the analyzed events, the flare area varied in a limited range throughout the events. The relative change of the area obtainted from linear fit for most of the flares did not reach level of 30\% (flares 1, 4, 8, 9, 10, 12, 13, 15). The total range of relative area changes was between 10\% and 61\%. In contrast, the flare temperature exhibited significant systematic variations increasing on average by factor 2.5. 
It should be noted that during the flare’s early phase—when the contrast ratio is sufficiently high—our estimated flare‐area values are quite reliable, with relative uncertainties of only about 10\%. In the late phase, however, as the contrast ratio drops, uncertainties can grow to exceed 100\% of the calculated area. Consequently, the adaptive rebinning procedure described in Subsection \ref{subsec:evol} is essential for stabilizing these estimates. By applying this rebinning, we obtain mean flare‐area values with relative uncertainties of 50\%. The most pronounced change in flare area occurred during the impulsive phase, when the flare fully emerges before reaching a stabilized state. This phase of the evolution is hard to control with the time cadence achieved in both our observations and in \cite{Howard_2020} data. In a contrast \cite{Hawley_2003} or \cite{Kowalski_2013} have reported significant variations in both flare area and temperature. The results for the analyzed flares may not necessarily apply to all flares. 
Flares No. 6 and 10 exhibit significant uncertainties in their peak temperature estimates, reaching up to 10$\,$000$\,$K. Since both the Evryscope and TESS bandpasses are located near the tail of the Planck curve, high temperatures result in very little energy being emitted within the TESS bandpass. This makes determining the temperature increasingly challenging and introduces significant uncertainties in the measured temperature values at higher temperatures for the flare No. 10. In the case of flare No. 6, the absence of TESS data, coupled with the flare's relatively low amplitude compared to the noise level in the $r'$ band significantly affected the accuracy of the derived temperature, leading to larger errors.

\subsection{Optical Thickness and Energy Considerations}
It is commonly assumed that the emission emitted by a flare in the continuum spectrum originates from a vertical narrow chromospheric condensation region \citep{Livshits_1981} and that the condensation maintains a constant kinetic temperature of approximately $10\,000\,$K.
In such a scenario, the optical continuum of superflares becomes optically thick causing the radiation to thermalize to the black-body at chromospheric condensation kinetic temperature.
However, this assumption is not always met, particularly for observed low-energy flares below a superflare threshold. This was demonstrated using the RHD simulations by \cite{Kowalski_2015, Kowalski_2017, Kowalski_2022} who explored the temperature, densities, and the optical-thickness evolution of the chromospheric condensation in solar and stellar flares. Flares with a chromospheric condensation layer characterized by a high electron density, $n_e > 5 \times 10^{14}\,$cm$^{-3}$, and peak black-body temperatures around 10$\,$000$\,$K or higher, exhibit optical spectra that closely match the Planck function and the condensation is optically thick in the Paschen continuum (e.g. \cite{Kowalski_2015}). 
The bolometric energies of flares 8, 10, 11, 12, 13, and 15 surpass the superflare energy threshold, with their peak temperatures exceeding $10\,000\,$K.
Assuming that these flares had optically thick continua, our energy estimations are well-suited for these events. The bolometric energies of flares 1, 2, 3, 4, 5, 6, 7, 9, and 14 were below the superflare energy threshold, suggesting they may not conform to the assumption of an optically thick black-body continuum spectrum. 
For these flares, an alternative method that includes the emission measure as an additional 
free parameter alongside the chromospheric condensation temperature and flare area may be 
necessary to accurately fit the observed flux enhancements using a simple cloud model for 
isothermal condensations \citep{simoes, Heinzel_2024}.

\begin{figure}[ht!]
    \centering
    \includegraphics[width=\columnwidth]{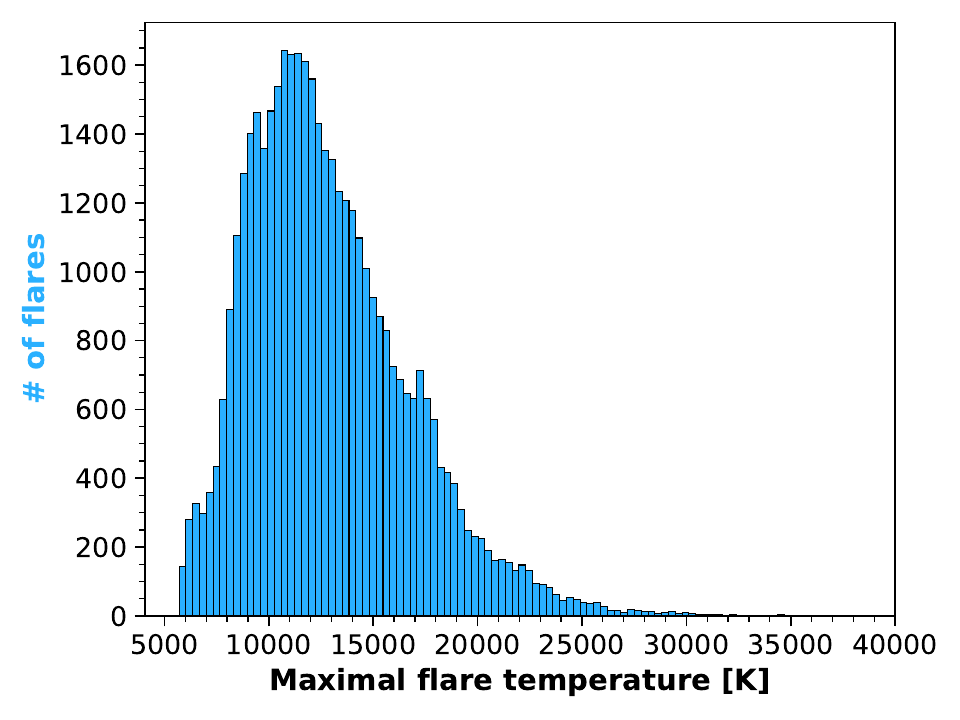}
    \caption{Distribution of the peak temperatures of 42$\,$257 flares on 268 stars, from \cite{Pietras_2022}, estimated using our semi-empirical grid.}
    \label{fig:histo}
\end{figure}

\subsection{Validation of Temperature Estimation Method}
The flare's area and temperature changes can be estimated using the peak black-body temperature during the event. To make this temperature estimation possible, we developed a semi-empirical grid based on the 14 out of 15 flares analyzed in this study. This grid establishes relationships among peak flare amplitude, peak black-body flare temperature, and effective temperature of the star. The grid is designed for use with main-sequence stars that have temperatures ranging from $2\,700\,$K to $4\,600\,$K, enabling the estimation of the peak flare temperature and its associated uncertainty. The grid provides a reasonably accurate approximation, with differences between the grid-estimated and observed values consistently falling within a 30\% margin. We tested our grid using three different spectral models: ATLAS9 \citep{kuruczspec}, PHOENIX, and TURBOSPECTRUM \citep{turbospectrum}. Regardless of which spectra were used to create the grid, we obtained very similar results for the maximum flare temperatures with differences up to 5\%.
We evaluated our methods for estimating temperature and area by analyzing the IF3 flare on YZ CMi, as described in \cite{Kowalski_2013}. This analysis also served to validate our grid against results obtained from spectroscopy. During the IF3 event, as estimated by \cite{Kowalski_2013}, the temperature of the flaring region rose from approximately $10\,000\,$K at the beginning of the flare to around $12\,100\,$K at its peak, and then decayed to approximately $7\,000\,$K. Simultaneously, the flare's area ranged between $0.2\%$ and $1.0\%$ of the stellar disk.
Using the $gri$ photometry included of this event \citep{kowalskivizier}, we determined the temperature evolution of IF3 with the method presented in Sect. \ref{subsec:ft}. Our estimation of the peak flare temperature was $13\,000 \pm 3\,200\,$K, which aligns well with the value of $\sim\!\!12\,100\,$K reported by \cite{Kowalski_2013}. While our estimated flare area showed minor variations, it could be approximated as a constant value of $0.70\% \pm 0.01\%$ of the stellar disk, which also agrees closely with the values reported by \cite{Kowalski_2013}.
It is important to note that we did not have {\it TESS} photometric data for this event, as {\it TESS} was launched in 2018, whereas the IF3 flare was observed in 2011. To estimate the peak temperature using our grid, we generated synthetic {\it TESS} data based on the analysis in \cite{Kowalski_2013} for IF3. This synthetic data was derived from the temperature and area obtained from spectra, which we consider the most reliable source. Using Eqs. \ref{eq8} and \ref{eq9}, along with the temperature and area evolution from \cite{Kowalski_2013} and the theoretical PHOENIX spectrum for YZ CMi, we estimated the flare amplitude to reach $\sim\!\!0.204$ in the normalized light curve of the star in {\it TESS} bandpass. This corresponds to a temperature of $14\,400 \pm 2\,600\,$K, as estimated using our grid. Additionally, we repeated the same process using the temperature and area evolution derived from $gri$ photometric estimations. In this case, the flare amplitude reached $\sim\!\!0.271$ in the normalized light curve in {\it TESS} bandpass, corresponding to a temperature of $15\,300 \pm 2\,800\,$K. These temperature estimates for the IF3 event were consistent within the error margins.

The key issue with the B-band is that it includes most higher-order Balmer lines rather than the Balmer jump. In addition to the line flux enhancement, as noted by \cite{castelli2004newgridsatlas9model, Kowalski_2013}, in this wavelength range, the Stark broadening of Balmer lines \citep{Peterson_1969, Donati_1985} creates a pseudo-continuum, which can lead to an apparent enhancement beyond what would be expected from an extension of the optical continuum. This depends on the density of the chromospheric flares. The last resolved Balmer line is a key indicator of density, according to the Inglis-Teller formula \citep{Inglis_1939}. The coresponding results of stellar flare simulations are shown in Figures 8, 10, and 11 of \cite{Kowalski_2015}. To assess the impact of Balmer lines on the B-band, we evaluated their contributions by analyzing the time-evolution spectra of the IF2, IF3 and IF7 flares on YZ CMi from \cite{Kowalski_2013}, applying the methodology described in our paper. We estimated black-body flare temperatures based on energy ratios in the B and pseudo-TESS bandpasses. While our pseudo-{\it TESS} bandpass does not fully cover the {\it TESS} range (extending to $\sim\!\!9200\,\AA$ instead of $\sim\!\!11\,000\,\AA$), it captures most of its sensitivity range, allowing a reasonable comparison. IF3 is classified as one of the most impulsive from \cite{Kowalski_2013} sample, whereas the flares IF2 and IF7 were classified as simple, classical flares with energies below superflare threshold. These three events were selected for their relatively high signal-to-noise ratio with wavelength range reaching up to $\sim\!\!9200\,\AA$. Our analysis showed that the ratio of energy emitted in spectrum of the flare compared to the black-body emission in the B-band varies between 2.5\% and 7.5\% throughout the IF3 flare event, and between approximately 2\% and 10\% throughout the IF2 and IF7 events, with the highest values occurring in the early decay phase, and lowest in the late decay phase. This leads to a possible overstimations by the value up to 10\%. Given the measurement uncertainties, this effect is negligible and does not significantly impact our results. Similarly, the temperature estimations using B-band and pseudo-{\it TESS} band-passes lead to similar temperature evolution values as estimated by \cite{Kowalski_2013} within the uncertainty range of up to approximately 10\% of the value. The broadening of high-order Balmer lines indeed contributes to the blue continuum, but our estimates indicate that this may not lead to a substantial overestimation of the temperature when using both B-band and {\it TESS} data.

\subsection{Flare Energy Estimates and Limitations}
Although our grid requires further refinement and the inclusion of additional flares, this initial version provides a valuable tool for approximating flare temperatures. 
In the next phase of analysis, we estimated the total energy of the IF3 event based on the temperature evolution derived from photometry, spectra, and Method II. Using Method II, we calculated a total energy of $8.3 \times 10^{33}\,$erg. From the temperature evolution derived from spectra, we obtained an energy of approximately $1.6 \times 10^{33}\,$erg, and from the temperature evolution derived from $gri$ photometry, we calculated an energy of approximately $2.5 \times 10^{33}\,$erg.

The bolometric energies of the 15 analyzed flares ranged from \flaree{2.0}{1.3}{31}$\,$erg to \flaree{1.2}{0.5}{35}$\,$erg based on the observational data. 
The total flare energy is directly proportional to the projected flare area on the star (Eq. \ref{eq:eq7}). However, without precise details about the flare's position on the stellar disk, the actual flare area could be significantly larger than the projected area. This discrepancy likely results in an underestimation of the flare energy. Consequently, our current energy estimates should be regarded as lower bounds.
The energy values differed by less than 1\% regardless of the area evolution was based on actual data or approximated as a constant mean value throughout the event. A similar outcome was observed when estimating bolometric energy using either a constant mean area or a minimally evolving area derived from interpolated values of the flare's peak temperature and light curve. 

In the case of the energy estimation method introduced by \cite{Shibayama_2013}, changes in temperature values can significantly affect energy estimates, although this effect is not always obvious. It arises from the anti-correlation between temperature values and the flare emission area. It should also be noted that this influence varies with the observational band used. For example, in the {\it TESS} band, when the flare peak temperature is around $10\,000\,$K, bolometric energy estimations depend in the black-body model as follows: $E\sim T^2$. However, at $20\,000\,$K, this dependence shifts to approximately $E\sim T^{2.5}$. In contrast, if the flare temperature is around 5500$\,$K, the flare energy remains largely unaffected by changes in temperature. All the above calculations were performed assuming an effective stellar temperature of 3300$\,$K, which is the average value for spectral type M.

Our analysis of 42$\,$257 peak black-body flare temperatures from 268 of the most active stars in the {\it TESS} data, with temperatures lower than 4600$\,$K, reveals that over 50\% of the flares have peak temperatures around 11$\,$100$\,\pm\,$2$\,$400$\,$K. Flares typically have a fast rise and a slower exponential decay, which means that during the majority of the flare, the temperature will be notably lower than the peak value (less than $10\,000\,$K). Consequently,  the Method II may overestimate flare energies in such cases. The flares No. 1 and No. 6 are good examples of this overestimation, as their peak temperatures are $5\,700\, \pm\, 450\,$K and $17\,500\, \pm\, 10\,000\,$K, respectively. However, temperatures above $10\,000\,$K are observed for only 90 seconds, or 5\% of the total duration for the flare 6, and the flare 1 does not reach $10\,000$K at all. Additionally, the areas estimated using  the Method II are lower for the flare 1 by a factor of 3, and higher the flare 6 by a factor of 15. This leads to an overestimation of flare energy by 300\% -- 400\% for both flares.

\subsection{Big Flare Syndrome and Possible Statistical Trends}
For stellar flares, studies by \cite{Pietras_2022, Pietras_2023} and others suggest that the standard solar flare model is applicable beyond the Sun. Similar physical processes likely take place on other stars, though on a much larger scale, implying that the BFS phenomenon may also be present.  The "Big Flare Syndrome" (BFS) describes a statistical trend in which all energetic flare phenomena, including soft and hard X-ray emissions, become more intense in larger flares, regardless of specific physical mechanisms \cite{Kahler_1982}. This pattern, well-documented in solar flares, is expected to occur on other stars as well. The relationship between white-light, hard, and soft X-ray emissions can be attributed to non-thermal electrons: those accelerated in the corona during magnetic reconnection deposit energy in the chromosphere through bremsstrahlung, producing hard X-ray emission at the footpoints. Simultaneously, these electrons heat the chromosphere, leading to evaporation that fills magnetic loops and generates soft X-ray emission. The peak temperatures of the observed flares, when compared to their mean areas throughout the event, show a positive correlation (with a probability of $p = 0.98$). This correlation suggests that, in stellar flares, higher flare energies correspond to larger flare areas observed in white light and to higher peak temperatures, much like in the solar case. Thus, Fig. \ref{fig:corelat} likely illustrates this BFS relationship, as slightly larger flare regions appear to be associated with higher peak temperatures. However, due to the limited sample size, further observations are necessary to explore this scenario in greater detail.
Based on the estimated flare areas, we conducted statistical tests to examine whether the analyzed events form two distinct groups. Using Student’s t-test, we compared the peak black-body temperatures of the flares, obtaining a p-value of approximately 0.09 and a t-statistic of about 2.15. While this does not provide strong evidence for a statistically significant difference at the 5\% level, it suggests a potential trend where larger flares may have higher peak temperatures.
For flare areas, the Mann-Whitney U test yielded a p-value of approximately 0.001, indicating a statistically significant difference between the two groups. We also investigated whether flare sizes correlate with stellar spectral types and found that larger flares predominantly occurred on earlier-type stars (M4V to K5V), while smaller flares were mostly observed on later-type stars (M5.5V to M2V).
Further investigation is needed to explore these trends more thoroughly.

\section{Conclusions}
In summary, our analysis leads to two key conclusions:
\begin{enumerate}
    \item We examined the temperature evolution of seven newly observed flares. Flare areas of these flares, and additional flares from \cite{Howard_2020}, remained constant within the 30\% (on the average) throughout their duration and can be treated as fixed. The issue of how big is the population of flares that do exhibit significant changes in their areas needs further investigations.
    \item We developed a semi-empirical grid for estimating flare temperatures in main-sequence stars, applicable for stellar effective temperatures up to 4600$\,$K, using {\it  TESS} data. This tool has the potential to enhance flare energy estimation for missions such as {\it TESS} and ESA's {\it PLATO}. Future work should prioritize obtaining additional high-quality observations of flares on F and G stars. Such data are essential to validate whether the flare characteristics observed in M and K dwarfs are similarly applicable to these higher-temperature stars. In this context, our established relationships between the effective temperature and the radius of main-sequence stars could prove instrumental in calibrating the grid for these additional flaring stars, ultimately enhancing the robustness of our methods.
\end{enumerate}
The findings underscore the importance of improving flare observation statistics through multicolor photometry and spectroscopy. Achieving this requires coordinated observation campaigns utilizing both ground-based and satellite platforms. Looking ahead, designing a future space mission with capabilities to observe in two or more photometric bands, similar to {\it TESS} or {\it PLATO}, would be a significant step forward.

\section{Acknowledgements}
This work was partially supported by the program "Excellence Initiative - Research University" for years $2020-2026$ for the University of Wroc\l{}aw, project No. BPIDUB.4610.96.2021.KG. Additional partial support was also provided by the grant of the Czech Funding Agency No.22-30516K and RVO:67985815. D. Marchev acknowledges project RACIO supported by the Ministry of Education and Science of Bulgaria (Bulgarian National Roadmap for Research Infrastructure). The authors appreciate the constructive comments and suggestions from the anonymous referees, which have been very helpful in improving the manuscript. The computations were performed using resources provided by Wroc\l{}aw Networking and Supercomputing Centre (https://wcss.pl), computational grant number 569. This paper includes data collected by the {\em TESS} mission. NASA's Science Mission Directorate provides funding for the {\em TESS} mission.

\bibliography{sample631}

\end{document}